\documentclass[aps,prl,twocolumn,groupedaddress]{revtex4}

\usepackage{epsfig}

\begin{document}

\title{Reply to Comment on ``Simple One-Dimensional Model of Heat Conduction 
which Obeys Fourier's Law''}

\author{P.L Garrido, P.I. Hurtado}

\affiliation{Departamento de E.M. y F\'{\i}sica de la Materia, Facultad de Ciencias, 
Universidad de Granada, 18071 Granada, Spain.}

\date{\today}


\maketitle

In this reply we answer a comment by A. Dhar on our Letter 
\cite{Pedro}. In that paper we studied thermal conductivity in a 
one-dimensional gas of hard point alternating-masses particles, demonstrating that Fourier's law holds 
both in its statical and dynamical aspects.

\begin{figure}[b]
\centerline{
\psfig{file=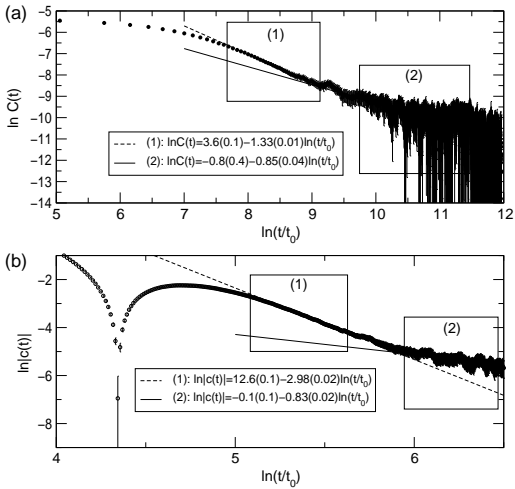,width=7.0cm,angle=-90}}
\caption{ \small $(a)$ $C(t)$ for the different masses system with $N=1000$. $(b)$ $c(t)$ for the equal 
masses system with $N=500$. The insets show the results of a power law fit for both regions ($1$) and ($2$). 
$t_0$ is the mean collision time.}
\label{uno}
\end{figure}

Dhar's main point is about the asymptotic behavior of the total energy current self 
correlation function $C(t)$. 
He fits $C(t) \sim t^{-0.83}$, implying an infinite 
thermal conductivity, while we obtained $C(t) \sim t^{-1.3}$ and thus a finite $\kappa$. 
In our opinion, Dhar's result is a consequence of autocorrelations 
due to finite size effects. In order to show this point, we plot
in fig. \ref{uno}.a $C(t)$ for a system size $N=1000$.
Here we can study two different regions: $(1)$ one for $\ln (t/t_0) \in [8,9]$, and $(2)$ other with 
$\ln (t/t_0)>10$. A power law fit to the first region yields $C(t) \sim t^{-1.3}$, while a fit to the 
second region yields $C(t) \sim t^{-0.88}$, very similar to Dhar's result. In fact, similar behavior has 
been measured by Savin et al\cite{Savin}. However, we think that only 
region $(1)$ corresponds to the infinite system asymptotic behavior. 
As an example of the previous statement, we can study the asymptotic behavior of the 
local energy current self correlation function $c(t)$ for the equal masses gas.
Following Jepsen \cite{Jepsen}, it can be shown analitically, after a lengthly calculation,
that $c(t) \sim t^{-3}$ for this system. We have measured $c(t)$\cite{Pedro} for 
equal masses in a finite system in the \textit{canonical ensemble} (\textit{not} in the zero-momentum 
ensemble, contrary to Dhar's comment). Fig. \ref{uno}.b shows $c(t)$ for $N=500$. 
It is remarkable that we can also define here two different regions: $(1)$ one for 
$\ln (t/t_0) \in [5.1,5.8]$, where a power law fit yields $|c(t)| \sim t^{-3}$, and $(2)$ one for 
$\ln (t/t_0) > 6$, where a power law fit yields $|c(t)| \sim t^{-0.83}$. 
We recover the theoretically predicted asymptotic bulk behavior in region $(1)$, 
while region $(2)$ should be due to finite size 
effects. Moreover, it is intringuing that the finite size time decay exponent 
($\sim 0.83$) is almost the same both in the different masses case and the equal masses one.
This fact points out the existence of an underlying common finite size 
mechanism, responsible of this spureous long time decay. 
In conclusion, coming back to the different masses case, we think that the above example indicates that
only region $(1)$ of fig. \ref{uno}.a represents the asymptotic bulk behavior. Hence, any conclusion about 
system's conductivity derived from region $(2)$ should be misleading.

Let's clarify now some other minor points raised in Dhar's comment. First, we {\it do not} observe linear
temperature profiles. They are linear in the central region and curved near the boundaries, which is consistent
with the finding of finite conductivity.\cite{Pedro}
Second, the validity of our deterministic heat bath and our local temperature measure has been 
carefully tested. 
On the other hand, it can be shown that $C(t) \sim Nc(t) + \sum_{i \neq l}c_{i,l}(t)$. 
Hence, for {\it regular} systems, where non-local time correlation functions $c_{i,l}(t)$ decay fast enough 
with distance, one expects a similar long time decay for both $C(t)$ and $c(t)$. However, there are 
{\it anomalous} systems, as the (non-ergodic) equal masses gas, for which $c_{i,l}(t)$ decays very slowly, 
or does not decay at all, and thus $C(t)$ and $c(t)$ behave completely different. 

In conclusion, we firmly confirm, after a global, consistent analysis of the problem, our previous 
results \cite{Pedro}, i.e. that our one--dimensional system
has a finite thermal conductivity in the Thermodynamic Limit, thus obeying Fourier's law.


\begin{thebibliography}{15}	
\bibitem{Pedro} P.L. Garrido et al, Phys. Rev. Lett. {\bf 86}, 5486 (2001).
\bibitem{Savin} A.V. Savin et al, Phys. Rev. Lett {\bf 88}, 154301-1 (2002).
\bibitem{Jepsen} D.W. Jepsen, J. Math. Phys. (N.Y.) {\bf 6}, 405 (1965).
\end{thebibliography}
\end{document}